\documentclass[twocolumn,showpacs,preprintnumbers,amsmath,amssymb]{revtex4}

\usepackage{graphicx}
\usepackage{dcolumn}
\usepackage{bm}

\begin{document}

\title{Consensus of self-driven agents with avoidance of collisions}
\author{Liqian Peng$^{1}$}
\author{Yang Zhao$^{1}$}
\author{Baomei Tian$^{1}$}
\author{Jue Zhang$^{1}$}
\author{Bing-Hong Wang$^{1,2}$}
\author{Hai-Tao Zhang$^{3,4}$}
\author{Tao Zhou$^{1,2,5}$}
\email{zhutou@ustc.edu}

\address
{$^{1}$Department of Modern Physics and Nonlinear Science  Center,
University of Science and Technology of China, Hefei 230026, P. R.
China \\ $^{2}$Research Center for Complex System Science,
University of Shanghai for Science and Technology, Shanghai 200093, P. R. China \\
$^{3}$ Department of Control Science and Technology, Huazhong
University of Science and Technology, Wuhan 430077, P.R. China \\
$^{4}$ Department of Engineering, University of Cambridge, Cambridge
CB2 1PZ, U.K. \\$^{5}$Department of Physics, University of Fribourg,
Chemin du Muse 3, CH-1700 Fribourg, Switzerland}

\date{\today}

\begin{abstract}
In recent years, many efforts have been addressed on collision
avoidance of collectively moving agents. In this paper, we propose a
modified version of the Vicsek model with adaptive speed, which can
guarantee the absence of collisions. However, this strategy leads to
an aggregated state with slowly moving agents. We therefore further
introduce a certain repulsion, which results in both faster
consensus and longer safe distance among agents, and thus provides a
powerful mechanism for collective motions in biological and
technological multi-agent systems.
\end{abstract}


\maketitle
\section{Introduction}
One of the most marvelous and ubiquitous phenomena in nature is
\emph{collective motion}, a kind of motion that can be observed at
almost every scale: from bird flocks and fish schools at a
macroscopic level to bacteria, individual cells and even molecular
motors at a microscopic level \cite{1,2,3,4,5,6,7,8,9}. Although in
most cases agents do not share global information and often travel
in the absence of leaders or external forces, collective motion may
still occur. Analogous behaviors are reported in engineering systems
also, such as groups of autonomous mobile robots and air vehicles
\cite{10,11,12,13,14,15,16} (see also a newly reported swarm model
that may connect granular materials and agent-based models
\cite{39}). In order to uncover the underlying mechanism leading to
the consensus of collective population, Vicsek \emph{et al}.
\cite{18} proposed a model with self-driven agents to mimic the
biological swarm, which displays a novel type of kinetic phase
transition. From then on, the nature of the nonequilibrium phase
transition of collective motion attracted more and more attentions
\cite{19,20,22,23,24,25}. Due to simplicity and efficiency, many
modified versions of the Vicsek model were proposed. For example,
some new methods with effective leadership were introduced
\cite{13,26,27}, and new moving protocols with adaptive speed to
accelerate the consensus were designed \cite{28,29}; meanwhile some
scholars have studied the consensus of collective motions via
low-cost communication \cite{30} and predictive mechanism
\cite{31,32,addPRE}, all of which can greatly enhance the global
convergence.

\begin{center}
\begin{figure}[!h]
\begin{center}
\includegraphics[scale=0.8]{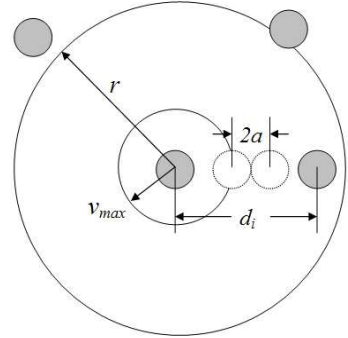}
\end{center}
\caption{Illustration of the current model with adaptive speed,
where $r$ denotes the horizon radius of an agent, $d_i$ denotes the
distance of the $i$th agent and its nearest neighbor, $v_{max}$
denotes the possibly maximal velocity, $a$ denotes the size of an
agent. Accordingly, $2a$ corresponds to the least distance of two
agents.}
\end{figure}
\end{center}

Recently, much attention has focused on how to keep distances among
agents. A common way is to introduce attraction and/or repulsion
\cite{13,14,39,Gregoire2003,21,33,34,35,36,37,38}. However, any
kinds of repulsions alone can not sufficiently avoid collision at
all times, because it is entirely possible that in a high-density
area, two agents are compelled to collide for the purpose of
avoidance of collision with a third agent. In this article, we
propose a swarm model with adaptive speed to completely eliminate
collisions. In a plane, each agent adjusts its direction as the
average direction of its neighbors while resets its speed according
to the minimal distance from its neighbors. The farther an agent is
away from its nearest neighbor, the higher speed it has. This
strategy can completely avoid collisions, however, it results in an
aggregated state where the agents move very slow in average.
Therefore, we further introduce a repulsion that can break down the
aggregation of agents, and thus sharply speeds up the global
convergence and enlarges the average distance among agents.

\section{model with adaptive speed}
We consider each agent as an inelastic ball with radius $a$, limited
in a square shaped cell of linear size $L$ with periodic boundary
conditions. Initially, each agent is randomly distributed in the
square, with moving direction randomly distributed in $[-\pi,\pi)$.
At each time step, the position of the $i$th agent is updated as:
\begin{equation}
\vec{x}_{i}(t+1)=\vec{x}_{i}(t)+\vec{v}_{i}(t),
\end{equation}
and its direction is updated as:
\begin{equation}
\theta_i(t+1)=\langle\theta_i(t)\rangle_r+\Delta\theta_i,
\end{equation}
where \(\Delta\theta_i\) denotes the thermal noise which is a random
number uniformly distributed in the interval $[-\eta, \eta]$ (In the
main context, we only consider the noise-free case, namely $\eta=0$.
A brief discussion about the effect of noise is presented in the
last section), \(\langle\theta_i(t)\rangle_r\) denotes the average
direction of the agents within the horizon radius $r$ of the $i$th
agent (including the $i$th agent itself), which reads:
\begin{equation}
\tan[\langle\theta_i(t)\rangle_r]=\langle
v_i\sin\theta_i(t)\rangle_r/\langle v_i\cos\theta_i(t)\rangle_r.
\end{equation}

\begin{center}
\begin{figure}[!htbp]
\scalebox{0.3}[0.3]{\includegraphics{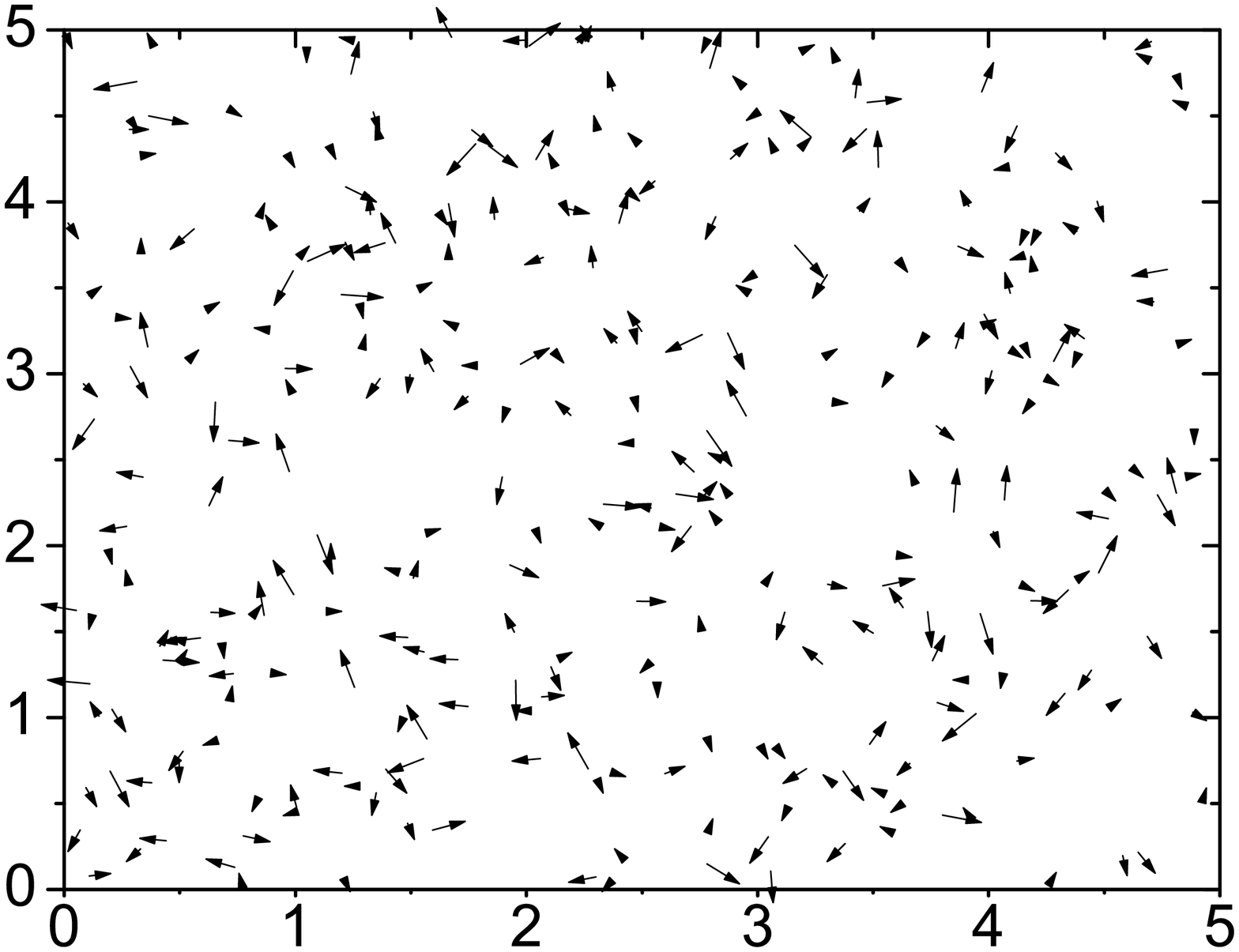}} {\begin{center}(a)
\end{center}}
\scalebox{0.3}[0.3]{\includegraphics{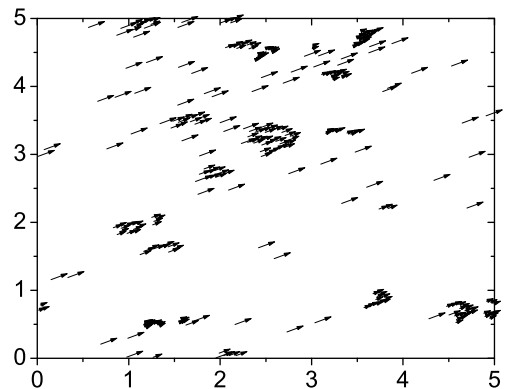}} {\begin{center}(b)
\end{center}}
\caption{Illustrations of locations and velocities in the initial
configuration (a), and at the 500th time step (b). The parameters
are set as $L$=5, $N$=300, $r$=1, $v_{max} $=0.03 and $a$=0.01. The
length and direction of an arrow represent the absolute value and
direction of the corresponding agent's velocity.}
\end{figure}
\end{center}
\begin{figure}[!htbp]
\includegraphics[scale=0.4]{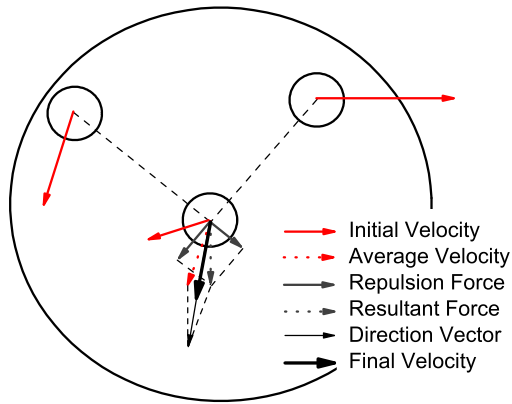}
\caption{(Color online) Illustration of the motion protocol with
repulsive effect.}
\end{figure}
\begin{figure}[!htbp]
\scalebox{0.3}[0.3]{\includegraphics{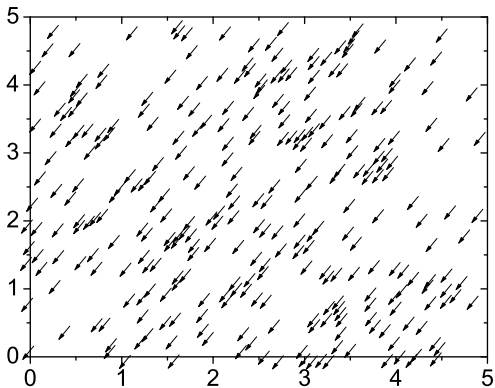}} \caption{The
distribution of positions and velocities at 500th time step in the
scattering model. The parameters are set as $L$=5, $N$=300, $r$=1,
$v_{max}$=0.03 and $a$=0.01. The length and direction of an arrow
represent the absolute value and direction of the corresponding
agent's velocity.}
\end{figure}

In natural swarms, the speed of each agent is alterable, that is,
agent may adjust not only its moving direction, but also its
absolute velocity. In the common sense, to avoid collisions with
other agents, an agent in a high-density group should adopt lower
speed. Taking urban traffic as an example, the speed of an
automobile is very low in the near-jammed situation, whereas it is
generally of high speed when sparse automobiles taking up the road.
Accordingly, we set the speed of the $i$th agent not more than
\(v_i=(d_i-2a)/2\), where \(d_i\) is the geographical distance
between two centers of the $i$th agent and its nearest neighbor (see
the illustration shown in Fig. 1). No matter how the $i$th agent and
its nearest neighbor choose their directions in the next time step,
the restriction can guarantee the distance between them no less than
$2a$ and therefore avoid collision. In fact, this restriction is not
only sufficient, but actually necessary. As the direction
\(\theta_i(t+1)\) of each agent in the next time step is determined
by the average direction within its own horizon radius, it is
impossible for an agent to know all the information of its
neighbors, especially the moving directions of its neighbors in the
next time step. Therefore, in order to avoid collision, it is
obliged to take into account the worst circumstance, that is, two
neighbors mutually approach. In this worst case, keeping the speed
of each agent (labeled by $i$) no more than $(d_i-2a)/2$ is the only
way guaranteeing the absence of collisions.

Accordingly, the absolute velocity of each agent is updated with the
following rule:
\begin{equation}
v_i (t + 1) = {\rm{Min}}\left( {v_{\max } ,\frac{{d_i- 2a}}{2}}
\right).
\end{equation}
Clearly, when the distance between an agent and its nearest neighbor
is longer than \(2v_{max}+2a\), its following speed can achieve the
maximum; otherwise, its speed is limited as $(d_i-2a)/2$.

Moreover, in order to quantify the consensus of moving directions,
an order parameter \cite{18} is introduced as:
\begin{equation}\label{va}
V_a  = \frac{{\left| {\sum\nolimits_{i = 1}^N {\vec{v}_i  } }
\right|}}{{\sum\nolimits_{i = 1}^N {v_i } }},\quad 0\leq V_a \leq 1,
\end{equation}
where \(v_i=|\vec{v}_{i}|\). A larger value of \(V_a\) indicates
better consensus. Since the speed in this model is no longer
constant, it is necessary to introduce another order parameter
\(V_b\) to evaluate the consensus of the absolute velocity, as:
\begin{equation}
V_b=\frac{\sqrt{\langle\Delta v^2\rangle}}{v}, \quad V_b\geq0,
\end{equation}
where \(v=\langle v_i\rangle\) is the average absolute velocity of
all the agents, and \(\Delta v^2\) is the variance of the absolute
velocity. Apparently, a smaller  \(V_b\) corresponds to better
consensus. Especially when \(V_b=0\), all agents share the same
speed.

\begin{figure}[!htbp]
\includegraphics[scale=0.3]{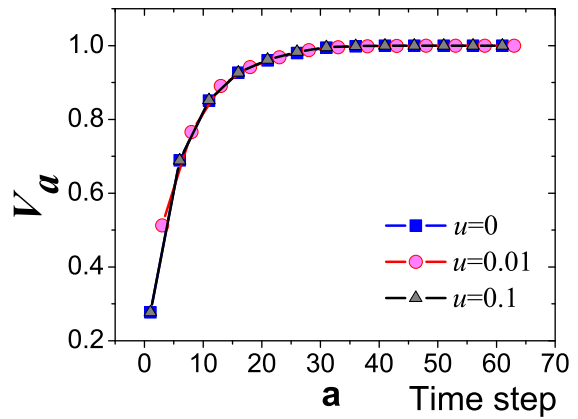}
\includegraphics[scale=0.3]{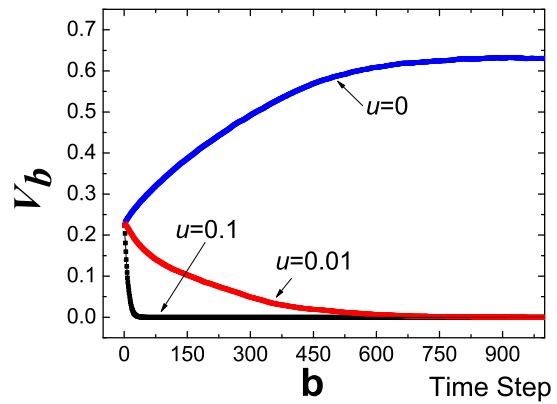}
\includegraphics[scale=0.3]{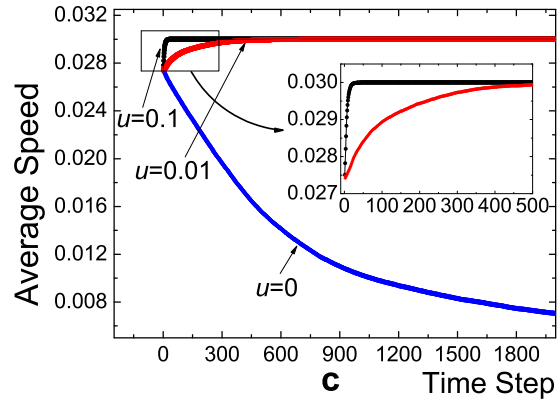}
\caption{ (Color online) $V_a$, $V_b$, and the average speed versus
time steps under different repulsion strengthes. The parameters are
set as $L$=5, $N$=300, $r$=1, $v_{max}=0.03$ and $a$=0.01. All the
data come form the average results of 500 independent runs.}
\end{figure}

Numerical results reveal that after the direction consensus, speed
still varies. Figures 2(a) and 2(b) respectively illustrate the
locations and velocities of all the agents in the initial
configuration and at the 500th time step. After a certain time
period from the beginning, the positions of agents are not uniformly
distributed and an aggregation phenomenon appears (see Fig. 2(b)).
Therein the average speed in a high-density area is much slower than
that in a low-density area. This aggregated state can be understood
as follows: agents in a high-density region agglomerate together and
mutually move in a low speed, thus they can seldom disperse apart.
Meanwhile they take up the way of their subsequent peers whose speed
is higher, making the high-density area congregate more agents, and
in turn achieving even higher density and slower speed (of course,
on the other hand, the density is limited by the size of agents,
$a$). Moreover, the motions of agents are similar to the laminar
flow in hydromechanics: when \(V_a\) gets close to 1, each agent is
moving along a line with the same direction and will never diverge
from its final track. Thus, different layers present various flowing
speeds.

In the current model, the nearest distance among agents in
high-density areas is very close to \(2a\), making the involved
agents move in a very low speed (close to 0); in the meantime the
nearest distances among agents in low-density areas are usually more
than \(2a+2v_{max}\), accordingly the involved agents can achieve a
high speed (close to \(v_{max} \)). Consequently, the absolute
velocities of all agents in the whole system can be in a high
diversity. Only in a low-density layer can the agents maintain
high-speed movement in a comparative long term. As a matter of fact,
the swarm never get speed consensus even with identical direction,
as shown later in Fig. 5(b). For the purpose of making all the
agents achieve the consensus with higher speed, it is necessary to
introduce a certain repulsion to avoid agglomeration. In addition,
denoting \(r_{ij}-2a\) as the safe distance between the $i$th and
the $j$th agents, where \(r_{ij}\) denotes the geographical distance
between the \(i\)th and the \(j\)th agents. In real applications of
unmanned air vehicles and auto-robots, the longer safe distance is
favorable. Therefore we hope a properly designed moving protocol
with repulsion could make the safe distance longer.

\section{scattering model}
Based on the strategy with adaptive speed mentioned above, in this
section, we introduce a repulsion to enlarge the safe distances
among agents. We assume: (1) the direction of the repulsion should
be along the line of two agents, and (2) the magnitude of the
repulsion should decrease with the increase of distance between two
agents. Moreover, as long as the distance between two agents is over
\(2v_{max}+2a\), no matter how they choose their directions and
speeds, collision will not occur in the following steps. Considering
this, the repulsion in our model should be a short-distance force
and work only when the distance is shorter than $r_{0}$
($r_{0}=2v_{max}+2a$). Accordingly, we set the form of repulsion
force as:
\begin{equation}
\vec{f}_{ij}=
    \left\{
    \begin{array}{cc}
    u\times\exp(-\frac{1}{1-r_{ij}/r_{0}})\times \frac{\vec{r}_{ij}}{r_{ij}}  & \mbox{$r_{ij}<r_{0}$}\\
    0 &\mbox{$r_{ij}\ge r_{0}$}
    \end{array}
    \right.
\end{equation}
where $u$ is a free parameter. Since the mass of an agent plays no
role in the present model, we suppose the repulsive effect (caused
by the repulsion) can directly affect the velocity vector in the
next time step (see Fig. 3 for an illustration).

\begin{figure}[!htbp]
\includegraphics[scale=0.3]{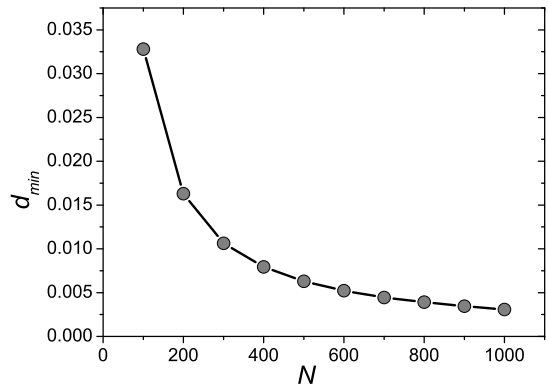}
\caption{Minimal geographical distance between pairs of agent in the
stable state of the standard Vicsek model, $d_{min}$, versus the
number of agents, $N$. Each data point is the average of 1000
independent runs. The restriction to avoid collisions with agent
size $a=0.01$ corresponds to $d_{min}>0.02$.}
\end{figure}

After introducing such a repulsion, the moving direction of each
agent is determined not only by the average direction within its
horizon radius, but also by the repulsive effect. The synthesis of
repulsive effect \(\vec{f}_i\) ($\vec{f}_i= \sum\nolimits_{j = 1}^N
{\vec f } _{ji}$, determined by Eq. (7)) and the average velocity
\(\vec{v}_i\) (whose direction and magnitude are respectively
determined by Eq. (3) and Eq. (4)) is set as the following moving
direction of the agent (see Fig. 3). On the other hand, the absolute
velocity should also follow the Eq. (4). Numerical simulations, as
shown in Fig. 4, indicate that this new protocol can effectively
scatter the aggregated agents (take Fig. 2(b) as an example for
comparison). Actually, under this protocol, each agent can hold a
certain distance (much longer than the system mentioned in Section
II) with its neighbors, and therefore achieves its maximal speed,
$v_{max}$.

We also investigate the effects of repulsion strength by adjusting
the parameter $u$. Figure 5(a) shows that the convergence of moving
direction is not sensitive to the repulsion strength. However, a
larger value of $u$ corresponds to a shorter time for the system  to
achieve the consensus of speed, as well as a higher average speed in
the steady state (see Fig. 5(b) and Fig. 5(c)). Considering Eq. (4),
the larger average speed actually implies that the average distance
between agents is longer. Note that, when $u=0$, $V_b$ can not
approach 0, and the average speed is very low.

\begin{figure}[!htbp]
\includegraphics[scale=0.3]{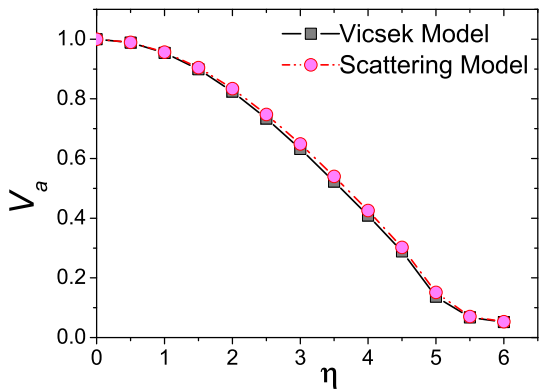}
\caption{ (Color online) Comparison of order parameter $V_a$ in the
Vicsek model and the scattering model under noisy environment. The
parameters are set as $L$=5, $N$=300, $r$=1 and $v_{max}$=0.03. In
scattering model, $u$=0.01 and $a$ =0.01. All the data come form the
average over 500 independent runs.}
\end{figure}

\begin{figure}[!htbp]
\includegraphics[scale=0.3]{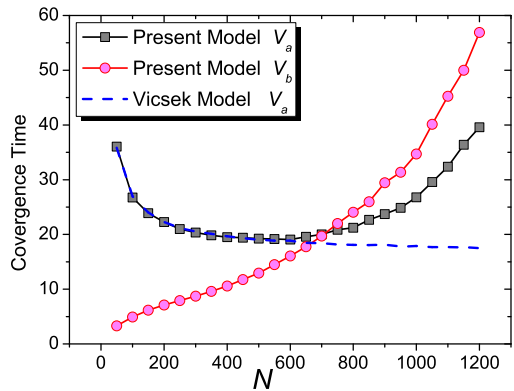}
\caption{(Color online) Comparison on convergence time between the
standard Vicsek model and the present model (i.e., the scattering
model) in the absence of noise. In the Vicsek model and the present
model, the convergence time for $V_a$ is defined as the required
time steps making $V_a$ larger than 0.99; while the convergence time
for $V_b$ is defined as the required time steps making $V_b$ smaller
than $10^{-3}$. Blue dash curve represent the simulation result for
the Vicsek model, while the black squares and red circles represent
the results for $V_a$ and $V_b$, respectively. The parameters are
set as $L$=5, $r$=1 and $v_{max}$=0.03. In the present model,
$u$=0.1 and $a$ =0.01. All the data come form the average over
$5\times 10^3$ independent runs.}
\end{figure}

\section {Conclusion and Discussion}
As long as we consider the sizes of agents, it is not only possible
but actually necessary to propose a protocol to avoid collisions
among them. Although the Vicsek model \cite{18} has achieved a great
success in mimicking the self-driven swarm, it cannot guarantee the
absence of collisions. We report in Fig. 6 a simple simulation of
the noise-free Vicsek model neglecting the sizes of agents. As the
growing of the population, in the stable state, the minimal
geographical distance between pairs of agents decreases quickly. If
the size of agent is set as $a=0.01$, then the minimal distance to
avoid collisions must be larger than $2a=0.02$. That is to say, the
standard Vicsek model can only hold less than 200 agents with size
0.01 in an $5\times 5$ square. In comparison, the current model with
adaptive speed can hold thousands of such agents.

However, numerical simulations showed an aggregation phenomenon in
the current model, which impedes the convergence of speed. To
overcome this blemish, we introduce a repulsion to scatter the
aggregated agents. The simulation results are exciting: Each agent
can hole a certain personal space; what is more, they can quickly
achieve speed consensus and move in a very high speed. Numerical
results also indicate that the stronger the repulsive effect is, the
less convergent time it takes to achieve the consensus. In section
II, we have already proved that even two neighbors mutually
approach, the adaptive strategy can still avoid possible collision.
Therefore, in any event, collision will never occur in the
scattering model.

Furthermore, it is well known that the thermal noise can also play a
significant role in determining the moving directions of agents.
Thus, we need to check whether our rule is robust in the presence of
noise. The numerical result indicates that even in the noisy
environment, in the stable state, the average distance and average
speed are both larger than those without the repulsion. The order
parameter for direction consensus of course decreases with the
increasing of noise strength, $\eta$, and it exhibits almost the
same trend as the standard Vicsek model (actually, it is a little
bit larger than the Vicsek model, see please the simulation result
shown in Fig. 7).

In the noise-free Vicsek model, given $r$ and $L$, the convergence
is faster with more agents (i.e., larger $N$) since they will have
more frequent communications in a denser circumstance. Actually, a
recent numerical study \cite{29} indicates that the convergence time
scales approximately as $(\texttt{ln}N)^{-1.3}$, that is, the larger
the population is, the shorter the convergence time is. In Fig. 8,
we report the simulation result on the convergence time in the
noise-free Vicsek model (see the blue dash curve), where the
threshold quantile is set as $V_a=0.99$. It decreases monotonously
with the increasing of $N$, in accordance with Ref. \cite{29}. In
contrast, in the present scattering model, more effort should be
taken to avoid collisions in the denser circumstance. Figure 8
compares the convergence time between the standard Vicsek model and
the scattering model in the absence of noise. One can find that in
the sparse circumstance, $N\leq 600$, the convergence times of the
Vicsek model and the scattering model are almost the same, while in
the denser range, the convergence time in the scattering model
quickly increases versus the slowly decreasing of that in the Vicsek
model. The convergence time for absolute velocity increases even
most quickly than that for moving direction. This result indicates a
limitation of the present model, namely it can not efficiently deal
with the systems with huge population. Accordingly, how to design an
efficient method to simultaneously guarantee the absence of
collisions and the quick convergence is still an open problem for
us. Anyway, in the case of $a=0.01$, the standard Vicsek model can
avoid the collisions only if the number of agents is less than or
about 100 (see Fig. 6), while the scattering model can hold about
600 agents with the same speed of convergence. We therefore believe
the scattering model can find applications in the design of motion
protocol for self-driven agents.

Some difficult yet important problems about the conservative model
remain to be further explored. For example, if the ahead ones of a
group of agents need not pay attention to the following ones (that
is, each agent only receives information in a sector ahead in its
moving direction rather than all the neighbors within its sight
radius \cite{Tian2008}), collisions may automatically disappear. If
the swarm needs shorter time to get convergence while avoiding the
collisions, it may indicate that the complete communication is not
always the most efficient manner while partial communication may be
better in some cases \cite{30,Tian2008}. In addition, the properties
of the phase transition induced by the noise (see, for example, in
Ref. \cite{21}, Gr\'egoire and Chat\'e showed that a swarm model
with repulsion as well as the minimal Vicsek model suffers a
discontinuous phase transition) remains an open issue. Though not
the focus in this article, it worths an detailed investigation in
the future.

\begin{acknowledgments}
This work is funded by the National Basic Research Project of China
(973 Program No.2006CB705500), the Specialized Research Fund for the
Doctoral Program of Higher Education of China under Grant No.
20060358065, the National Science Fund for Fostering Talents in
Basic Science under J0630319, and the National Natural Science
Foundation of China under Grant No. 10532060. H.T.Z. acknowledges
the support of the National Natural Science Foundation of China
(NNSFC) under Grant No. 60704041, and the Specialized Research Fund
for the Doctoral Program of Higher Education of China under Grant
No. 20070487090. T.Z. acknowledges the National Natural Science
Foundation of China under Grant No. 10635040.

\end{acknowledgments}

\end{document}